\documentclass{PoS}

\title{String tension scaling in models of the confined phase}

\ShortTitle{String tension scaling in models of the confined phase}

\author{Peter N. Meisinger and \speaker{Michael C. Ogilvie}\\
        Washington University Dept. of Physics, St. Louis, MO 63130 USA\\
        E-mail: \email{pnm@wuphys.wustl.edu},
        \email{mco@wuphys.wustl.edu}}

\abstract{We introduce a D-dimensional Hamiltonian formalism for the study of
Polyakov loop models of finite temperature gauge theories in D+1 dimensions.
Polyakov loop string tensions are obtained from energy eigenstates
of the Hamiltonian. For D=1, the gauge theory reduces to quantum mechanics
on the gauge group; for D>1, the Hamiltonian includes hopping terms
that link sites on the transverse lattice. The deconfined phase is
associated with a ground state which breaks Z(N) symmetry, and Svetitsky-Yaffe
critical universality emerges naturally for D>1. A minimal model is
proposed which naturally reproduces approximate Casimir scaling for
a range of couplings. Different classes of potentials lead to different
pictures of how confinement is realized. Such potential energy terms
also modify string tension scaling laws, as we demonstrate using two
potentials: one representing the perturbative thermal contributions
from gluons, and the other arising from magnetic monopoles in certain
confining supersymmetric theories.}

\FullConference{XXIVth International Symposium on Lattice Field Theory\\
                July 23-28, 2006\\
                Tucson, Arizona, USA}

\begin{document}

\section{Introduction}

There is a substantial body of evidence from lattice simulations
\cite{Deldar:1999vi,Bali:2000un,Lucini:2001nv,Lucini:2002wg,DelDebbio:2001sj,DelDebbio:2003tk,Lucini:2004my}
that string tensions in confining gauge theories
obey scaling laws.
One strong possibility is Casimir scaling,
where the ratio of string tensions obey
${\sigma_{R'}}/{\sigma_{R}}={C_{2R'}}/{C_{2R}}$
with $C_{2R}$ the quadratic Casimir for the
representation $R$.
Another possible scaling law, suggested by string
theory, 
is sine-law scaling,
in which the lowest string tension in each $N$-ality sector obeys
${\sigma_{k}}/{\sigma_{1}}=
\sin\left( k\pi/N\right)/\sin\left( \pi/N \right)$
 \cite{Douglas:1995nw,Hanany:1997hr}.
We will explore
string tension scaling using a class of Hamiltonian effective models for Polyakov loops
obtained from underlying $D+1$-dimensional gauge theories
using the same type of arguments
which lead to the Euclidean effective
action for Polyakov loops \cite{Svetitsky:1982gs}.
These models are
defined on a $D$-dimensional space
with one continuous dimension and $D-1$ discrete directions. 
The general form of the Hamiltonian may be written as
\begin{equation}
H=\sum_{j}\left[\kappa C_{2j}+V_{j}\right]-\frac{J_{F}}{2}\sum_{\left\langle jk\right\rangle }\left[\chi_{Fj}\chi_{Fk}^{*}+\chi_{Fj}^{*}\chi_{Fk}\right]\end{equation}
The first term is the kinetic term for each site $j$, given by the
Casimir operator. Higher-order terms may also be present.
The second term is the potential on each site, $V_{j}$, a $Z(N)$-invariant
class function of the group. The third term is a hopping term between
nearest-neighbor sites, and generalizes to include other representations
and longer hops. $H$ acts on wave functionals which are class functions 
$\Psi\left[P\right]$ of Polyakov loops. 
The Hilbert space is spanned by products of characters,
with Haar measure providing the natual inner product. The integration
over each Polyakov loop need only be taken over the maximal Abelian
subalgebra, so that each Polyakov loop $P_{j}$may be represented
as $P_{j}=diag\left[e^{i\theta_{1}}..e^{i\theta_{N}}\right]$.
The ground
state energy density $E_{0}$ in the Hamiltonian formalism is related
to the free energy density $f$ and the pressure $p$ via $\beta f=-\beta p=E_{0}$.

A central question we address is naturalness: what
class of potentials lead to Casimir scaling or similar behavior?
Many potentials have regions of parameter space which tend to localize
wave functions around $N$ degenerate minima related by $Z(N)$ symmetry.
In such cases, $Z(N)$
symmetry can be maintained in the confined phase by tunneling. If
we have an approximate wave function $\Phi_{0}\left(P\right)$ which
is localized near one of the degenerate minima, we can construct
a set of $N$ wave functions as \begin{equation}
\Psi_{k}(P)=\frac{1}{\sqrt{N}}\sum_{j=0}^{N-1}\left[e^{2\pi ijk/N}\Phi_{0}\left(e^{-2\pi ijk/N}P\right)\right]\end{equation}
It is easy to show that the string tension splitting in this case
is determined by tunneling, giving a scaling law of the form
${\sigma_{k}}/{\sigma_{1}}=
\sin^2\left( k\pi/N\right)/\sin^2\left( \pi/N \right)$.
\begin{figure}
\centering
\includegraphics[width=.5\textwidth]{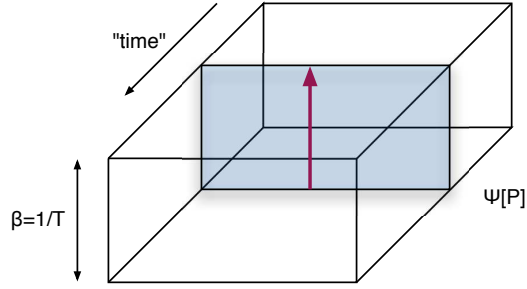}
\caption{Lattice geometry for Hamiltonian formalism.}
\label{fig:Lat06Formalism1}
\end{figure}
Another, complementary limit localizes the wave function around the
unique set of eigenvalues left invariant by Z(N) symmetry. For $SU(3)$,
this gives the matrix $P_{0}=diag\left[1,e^{2\pi i/3},e^{4\pi i/3}\right]$.
For any representation $R$ with non-zero $N$-ality,
$\chi_{R}\left(P_{0}\right)=0$,
so a configuration of Polyakov loops of the form $g(x)P_{0}g^{-1}(x)$
can be said to confine. 
These limiting behaviors represent the extremes of the potentials
we discuss below.

\section{Mean Field Theory}

In a Hamiltonian formalism, mean field theory is equivalent to the
Hartree approximation. We assume that the ground wave function is
a product of uncorrelated site wave functions:
$\Psi\left[P\right]=\prod\psi_{j}\left(P_{j}\right)$.
Minimizing
the energy and requiring self-consistency gives the Hamiltonian $H_{j}$
for the site $j$:\begin{equation}
H_{j}=\kappa C_{2j}+V_{j}-\frac{J_{F}d}{2}\left[\chi_{Fj}\left\langle \chi_{F}^{*}\right\rangle +\chi_{Fj}^{*}\left\langle \chi_{F}\right\rangle \right]\end{equation}
In the confined phase, $Z(N)$ symmetry requires $\left\langle \chi_{F}\right\rangle =0$,
and we are left with a Hamiltonian for $Z(N)$-invariant quantum mechniacs
on the group manifold:\begin{equation}
\left[\kappa C_{2j}+V_{j}\right]\psi_{jn}\left(P_{j}\right)=E_{n}\psi_{jn}\left(P_{j}\right)\end{equation}
As a first approximation, we associate the eigenvalues with string tensions
via $E_{n}-E_{0}=\sigma_{n}/T$.

The Hamiltonian formalism is capable of reproducing the behavior of the
deconfinement phase transition revealed by familiar models
based on an effective action \cite{Svetitsky:1982gs}.
In the case of $SU(2)$, assume for simplicity
a potential $V_j = -h_A \chi_{Aj}$ and
a trial wave function of the form
$\psi_{0j}=\sqrt{1-A^{2}-B^{2}}+A\chi_{Aj}+B\chi_{Fj}$
where $A$ and $B$ can be taken to be real.
The order parameter for the deconfinement transition is
the expected value of the Polyakov loop in the fundamental representation
$\left\langle \chi_{F}\right\rangle \equiv m\left(A,B\right)=2B\left(A+\sqrt{1-A^{2}-B^{2}}\right)$.
The ground state energy density
is given by the minimum of
\begin{equation}
E_{var}=\kappa\left(C_{A}A^{2}+C_{F}B^{2}\right)-\left(D-1\right)J_{F}m^{2}\left(A,B\right)-h_{A}\left(A^{2}+2A+B^{2}\right)\end{equation}
The low-temperature, confined phase occurs when the term in $E_{var}$
proportional to $\kappa$ dominates, and $E_{var}$is minimized when
$B=0$. If $h_{A}$ is non-zero, $A$ will be non-zero, and $\left\langle \chi_{A}\right\rangle $
will have a non-zero expectation value at all temperatures. As the
temperature grows, the hopping parameter $J_{F}$ grows
as well. When the coefficient of $B^{2}$ in $E_{var}$ becomes negative,
there is a second-order phase transition to the deconfined phase where
$B\neq0$ and $Z(2)$ symmetry is spontaneously broken. Note that
the $h_{A}$ potential term causes the critical temperature to change,
but not the order of the transition.

In the case of $SU(3)$,
we take a trial wave function of the
form $\psi_{0j}=\sqrt{1-A^{2}-2B^{*}B}+A\chi_{Aj}+B^{*}\chi_{Fj}+B\chi_{Fj}^{*}$
and the order parameter is given by
$\left\langle \chi_{Fj}\right\rangle \equiv m\left(A,B\right)=2B\left(A+\sqrt{1-A^{2}-2B^{*}B}\right)+\left(B^{*}\right)^{2}$.
The variational estimate of the ground state energy density is given by
 \begin{equation}
E_{var}=\kappa\left(C_{A}A^{2}+C_{F}B^{2}\right)-\left(D-1\right)J_{F}\left|m\left(A,B\right)\right|^{2}-h_{A}\left(2A^{2}+2A+2B^{*}B\right)\end{equation}
The new feature in $SU(3)$is the occurrence of a term proportional
to $J_{F}\left(B^{3}+B^{*3}\right)$. Such a term is permitted by
$Z(3)$ symmetry and makes the deconfining transition first order.

\begin{figure}
\centering
\includegraphics[width=.7\textwidth]{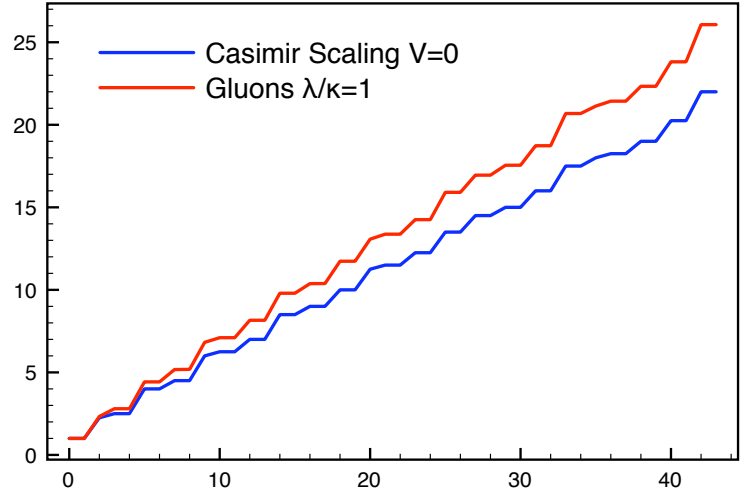}
\caption{String tension ratios for thermal gluon potential.}
\label{fig:CasimirRatiosFTb1}
\end{figure}
\begin{figure}
\centering
\includegraphics[width=.7\textwidth]{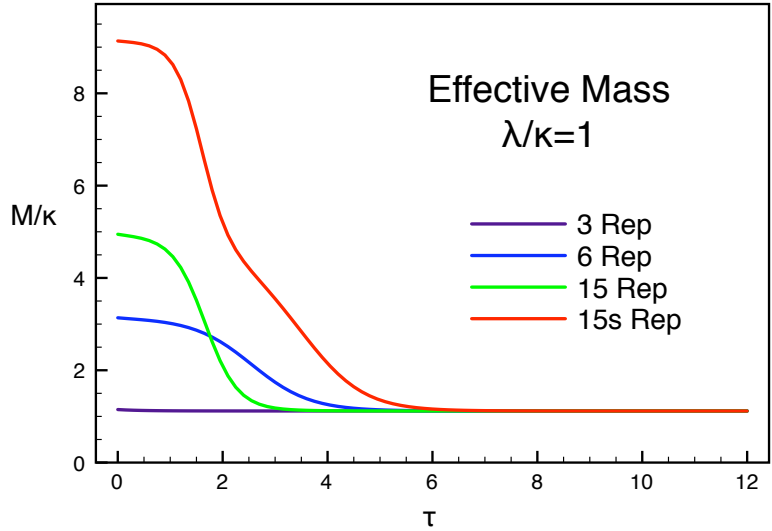}
\caption{Effective Masses for thermal gluon potential.}
\label{fig:PropFTLambda1}
\end{figure}

\section{Potentials}

\begin{figure}
\centering
\includegraphics[width=.7\textwidth]{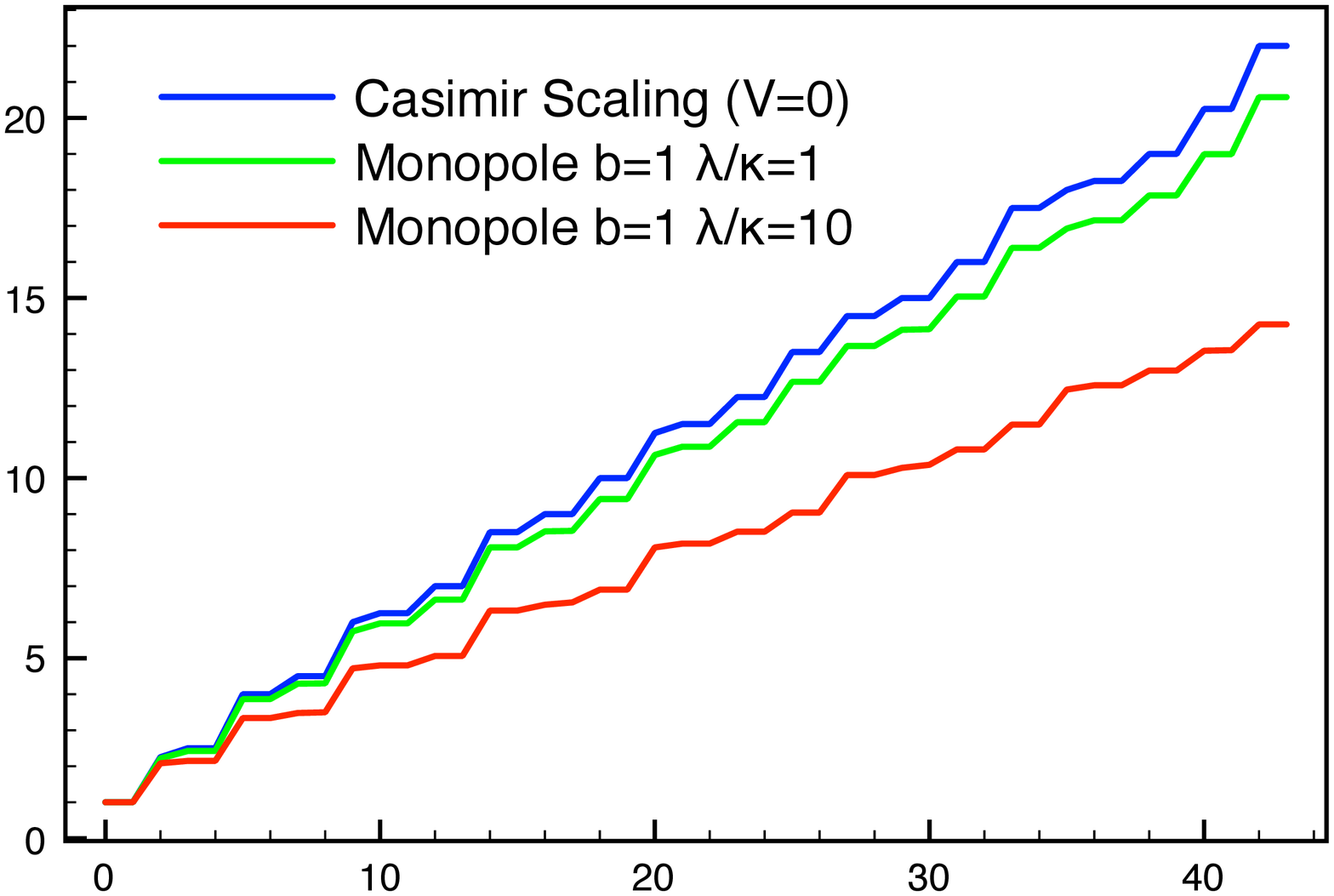}
\caption{String tension ratios for $W$ at $B=1$.}
\label{fig:CasimirRatiosMonopoleB1}
\end{figure}
\begin{figure}
\centering
\includegraphics[width=.7\textwidth]{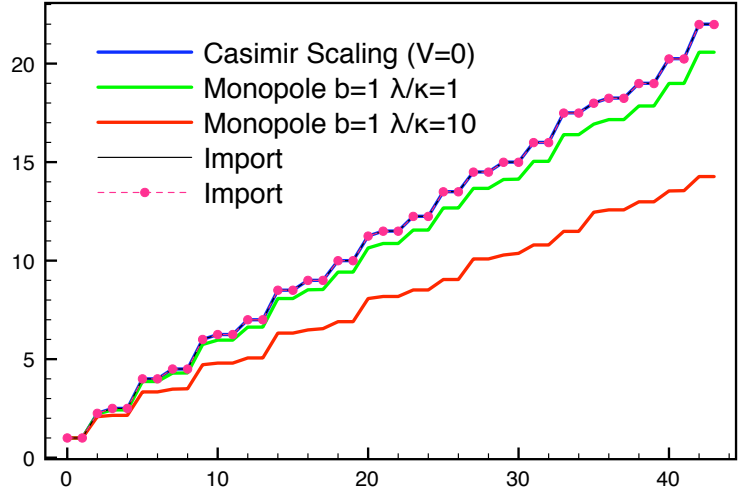}
\caption{String tension ratios for $W$ at $B=10$.}
\label{fig:CasimirRatiosMonopoleB10}
\end{figure}

Particle loop effects contribute
to the potential terms in the effective Hamiltonian.
The one-loop contribution of the massless gauge bosons to $V$ is
\begin{equation}
-\lambda\sum_{n=1}^{\infty}\frac{1}{n^{4}}\chi_{A}\left(P^{n}\right)\end{equation}
with $\lambda = 2T^4/{\pi^2}$.
This potential is minimized at
the elements of $Z(N)$,
and yields the familiar black-body formula for the free energy density
at those points.
We have calculated the string tensions
for this potential by 
diagonalizing the 
quantum-mechanical $SU(3)$ Hamiltonian on a finite basis,
using as a basis all
representations up to eight columns wide in the Young tableaux.
In figure 2 we compare
the sorted string tension ratios $\sigma_R/\sigma_F$
for $\lambda/\kappa=1$ with Casimir scaling.
Although there are
systematic deviations, they are most apparent
for larger representations.
We can define an effective mass
associated  with the exponential decay of
each representation as
a function of the scaled time 
$\tau = (E_1-E_0)t$.
Figure 3 shows
substantial mixing between representations, as evidenced by the rapid
approach of effective masses to the lowest mass.

\begin{figure}
\centering
\includegraphics[width=.7\textwidth]{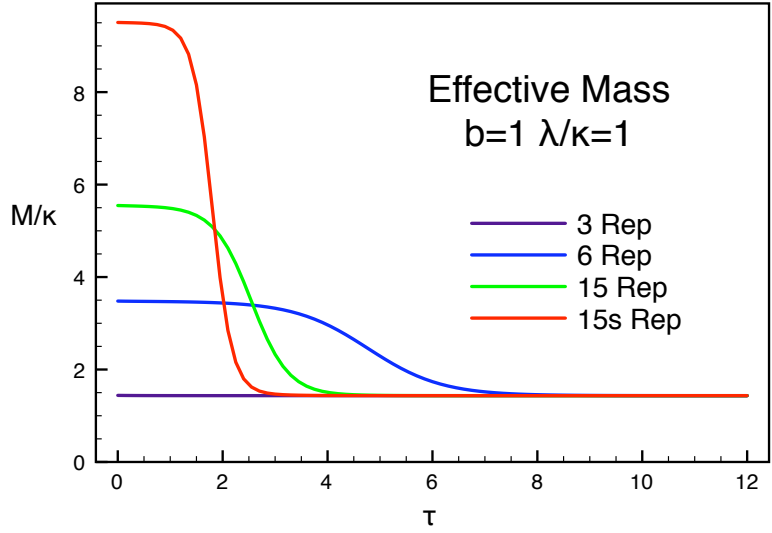}
\caption{Effective Mass Ratios at $b=1$.}\label{fig:PropB1Lam1}
\end{figure}
\begin{figure}
\centering
\includegraphics[width=.7\textwidth]{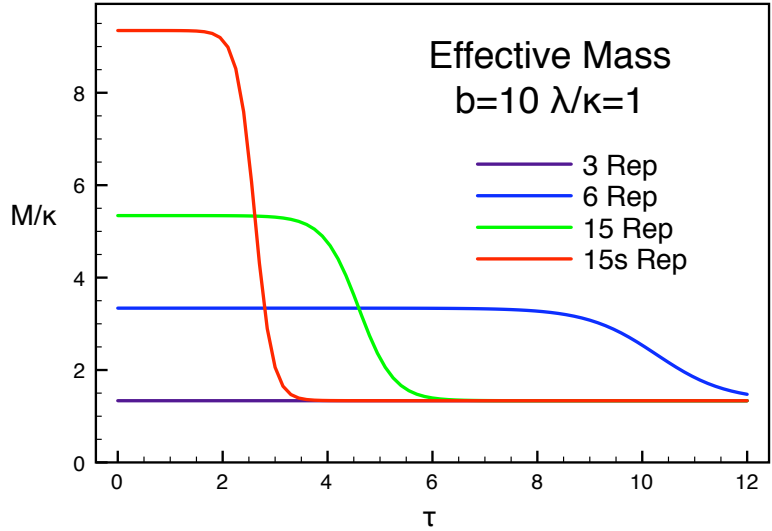}
\caption{Effective Mass Ratios at $b=10$}\label{fig:PropB10Lam1}
\end{figure}

Topological excitations also contribute to the potential.
For example, the one-loop calculation in $SU(2)$
of caloron contributions to the functional integral
in $SU(2)$ at finite temperature 
indicates an instability of the deconfined plasma phase
at sufficiently low temperature \cite{Diakonov:2004jn}.
In certain supersymmetric $SU(N)$ gauge theories on $S^{1}\times R^{3},$
confinement can be understoood as arising 
from magnetic monopole effects \cite{Diakonov:2004jn,Davies:1999uw}.
The superpotential $W$, regarded as a function of the Polyakov
loop eigenvalues, is the affine Toda potential,
 and is minimized at $P_0$. In the case of $SU(3)$, the
superpotential can be written in the form
\begin{equation}
W=\lambda\left[e^{-b\left(\theta_{1}-\theta_{2}\right)}+e^{-b\left(\theta_{2}-\theta_{3}\right)}+e^{-b\left(2\pi-\theta_{1}+\theta_{3}\right)}\right]\end{equation}
where $b=4\pi/g^{2}$. The eigenvalues are taken to lie in the
first Weyl chamber, where the eigenvalues are ordered
as $\theta_{1}\geq\theta_{2}\geq\theta_{3}$. 
We will use $W$ as an example of a potential that is minimized
at $P_0$.

Figure 4 shows that the deviation from Casimir scaling
increases with the strength of the potential $\lambda$
as expected.
In comparison, figure 5 indicates that at fixed $\lambda$,
Casimir scaling is recovered in the limit of large $b$.
As $b$ increases, $V$ is effectively zero except near
the boundary between Weyl chambers, where two or more
Polyakov loop eigenvalues coincide.
Because Haar measure leads to eigenvalue repulsion,
the wave function naturally vanishes when two
eigenvalues coincide, and for
large $b$, $W$ has
little effect on the wave functions.

The effective masses, shown in figures 6 and 7, show
longer plateaus at larger $b$. This is particularly noticeable
for smaller representations. This is consistent with
the wave functions for large $b$ approaching the $V=0$
wave functions. Thus increasing $b$ decreases mixing
between representations of the same $N$-ality.

\section*{Acknowledgments}
MCO gratefully acknowledges  the support of the U.S. Dept. of Energy.



\begin{thebibliography}{99}

\bibitem{Deldar:1999vi}
  S.~Deldar,
  %
  Phys.\ Rev.\ D {\bf 62}, 034509 (2000)
  [arXiv:hep-lat/9911008].

\bibitem{Bali:2000un}
  G.~S.~Bali,
  %
  Phys.\ Rev.\ D {\bf 62}, 114503 (2000)
  [arXiv:hep-lat/0006022].

\bibitem{Lucini:2001nv}
  B.~Lucini and M.~Teper,
  %
  Phys.\ Rev.\ D {\bf 64}, 105019 (2001)
  [arXiv:hep-lat/0107007].

\bibitem{Lucini:2002wg}
  B.~Lucini and M.~Teper,
  Phys.\ Rev.\ D {\bf 66}, 097502 (2002)
  [arXiv:hep-lat/0206027].
  
\bibitem{DelDebbio:2001sj}
  L.~Del Debbio, H.~Panagopoulos, P.~Rossi and E.~Vicari,
  JHEP {\bf 0201}, 009 (2002)
  [arXiv:hep-th/0111090].
  
\bibitem{DelDebbio:2003tk}
  L.~Del Debbio, H.~Panagopoulos and E.~Vicari,
  %
  JHEP {\bf 0309}, 034 (2003)
  [arXiv:hep-lat/0308012].

\bibitem{Lucini:2004my}
  B.~Lucini, M.~Teper and U.~Wenger,
   improved operators,''
  %
  JHEP {\bf 0406}, 012 (2004)
  [arXiv:hep-lat/0404008].
    
\bibitem{Douglas:1995nw}
  M.~R.~Douglas and S.~H.~Shenker,
   ``Dynamics of SU(N) supersymmetric gauge theory,''
  %
  Nucl.\ Phys.\ B {\bf 447}, 271 (1995)
  [arXiv:hep-th/9503163].
  
\bibitem{Hanany:1997hr}
  A.~Hanany, M.~J.~Strassler and A.~Zaffaroni,
  %
  Nucl.\ Phys.\ B {\bf 513}, 87 (1998)
  [arXiv:hep-th/9707244].

\bibitem{Svetitsky:1982gs}
  B.~Svetitsky and L.~G.~Yaffe,
  Nucl.\ Phys.\ B {\bf 210}, 423 (1982).

\bibitem{Diakonov:2004jn}
  D.~Diakonov, N.~Gromov, V.~Petrov and S.~Slizovskiy,
  Phys.\ Rev.\ D {\bf 70}, 036003 (2004)
  [arXiv:hep-th/0404042].

\bibitem{Davies:1999uw}
  N.~M.~Davies, T.~J.~Hollowood, V.~V.~Khoze and M.~P.~Mattis,
  Nucl.\ Phys.\ B {\bf 559}, 123 (1999)
  [arXiv:hep-th/9905015].

\bibitem{Davies:2000nw}
  N.~M.~Davies, T.~J.~Hollowood and V.~V.~Khoze,
  J.\ Math.\ Phys.\  {\bf 44}, 3640 (2003)
  [arXiv:hep-th/0006011].

   
\end{thebibliography}
\end{document}